\documentclass[aps,twocolumn,superscriptaddress,amsmath,amssymb]{revtex4-1}
\usepackage{graphicx}  
\usepackage{dcolumn}   
\usepackage{bm}        
\usepackage{verbatim}   
\usepackage{color} 
\usepackage{physics}

\begin{document}

\affiliation{Univ Insubria, Dipartimento Sci \& Alta Tecnol, Via Valleggio 11,
I-22100 Como, Italy} \affiliation{Linnaeus University,Department of Physics and
Electrical Engineering, 392 31 Kalmar, Sweden} \affiliation{CNR-IMM, sede Agrate
Brianza, via Olivetti 2, I-20864 Agrate Brianza, Italy}

\title{Impurity-induced topological phase transitions in $Cd_3As_2$ and $Na_3Bi$
Dirac semimetals}	

\author{A. Rancati} \affiliation{Univ Insubria, Dipartimento Sci \& Alta Tecnol,
Via Valleggio 11, I-22100 Como, Italy} \affiliation{CNR-IMM, sede Agrate
Brianza, via Olivetti 2, I-20864 Agrate Brianza, Italy} \author{N. Pournaghavi}
\affiliation{Linnaeus University,Department of Physics and Electrical
Engineering, 392 31 Kalmar, Sweden} \author{M. F. Islam} \affiliation{Linnaeus
University,Department of Physics and Electrical Engineering, 392 31 Kalmar,
Sweden} \author{A. Debernardi} \affiliation{CNR-IMM, sede Agrate Brianza, via
Olivetti 2, I-20864 Agrate Brianza, Italy} \author{C.M. Canali}
\affiliation{Linnaeus University,Department of Physics and Electrical
Engineering, 392 31 Kalmar, Sweden}

\date{\today}

\begin{abstract} 
Using first-principles density functional
theory calculations, combined with a topological analysis, we have investigated
the electronic properties of $Cd_3As_2$ and $Na_3Bi$ Dirac topological
semimetals doped with
non-magnetic and magnetic impurities. 
Our systematic analysis shows that the selective breaking of
the inversion, rotational and time-reversal symmetry,
controlled by specific choices of the impurity doping, induces
phase transitions from the original  Dirac semimetal to a variety of topological phases such as, 
topological insulator, trivial
semimetal, non-magnetic and magnetic Weyl semimetal, and  Chern insulator.
The Dirac semimetal phase can exist only if the rotational symmetry $C_n$ with $n > 2$ is maintained.
One particularly interesting phase 
emerging in doped $Cd_3As_2$ is a coexisting Dirac-Weyl phase, which occurs when only inversion
symmetry is broken while time-reversal symmetry and rotational symmetry are both
preserved. 
To further characterize the low-energy excitations of this
phase,
we have complemented our density functional results with a
continuum four-band $k\cdot p$ model, which indeed displays nodal points of both Dirac and Weyl type.
The coexisting phase appears as a transition point between two
topologically distinct Dirac phases, but may also survive 
in a small region of parameter space controlled by
external strain. 
\end{abstract}

\maketitle

\section{Introduction} \label{intro}

Topological Dirac materials are a class of advanced materials characterized by
electronic excitations with a linear dispersion about some nodal points close to
the Fermi level\cite{Hasan2010,Yan2012,Yan2017,Hasan2017,Chang2018}. 
Since the theoretical
prediction\cite{Kane2005,Bernevig2006} and the subsequent experimental
discovery\cite{Konig2007} of time-reversal-invariant three-dimensional (3D) topological
insulators (TIs), displaying two-dimensional (2D) surface states 
with a spin-momentum-locked linear dispersion around nodal Dirac points (DPs),
there has been an intense
search for materials that host 3D Dirac fermions with linear dispersion in all
three momentum directions.
By using theoretical models, the existence of such gapless
nodes in 3D bulk system was predicted by Murakami as a transition point between
the quantum spin Hall and the insulator phases\cite{Murakami2007}. In systems
possessing both inversion symmetry (IS)  and time-reversal symmetry (TRS), this
transition results in four-fold degenerate 3D Dirac nodes. For systems in which
IS or TRS is broken, the gapless nodes are two-fold degenerate nodes known as
Weyl points (WPs). Since the nodal points occur at around the Fermi energy,
these materials are known as Dirac semimetals (DSMs) and Weyl semimetals (WSMs),
respectively. Materials hosting topologically protected 
Dirac or Weyl fermions as elementary
excitations have the potential to revolutionize
low-energy high-performance spin-electronics. 
For a detailed review of the properties of
topological semimetals, see
Ref.~\onlinecite{Armitage2018}.        

The search for stable 3D DSMs in realistic systems experienced a breakthrough
with the studies of $Na_3Bi$ and $Cd_3As_2$  compounds. Using density functional
theory (DFT) methods, it was shown theoretically that a pair of stable 3D DPs,
protected by the rotational symmetry of the space groups of these crystals,
exists on the $k_z$ axis\cite{Wang2012,Wang2013}. Eventually, by employing
angle-resolved photoemission spectroscopy, these Dirac nodes were observed in
$Na_3Bi$\cite{Borisenko2014,Liu2014_1} and
$Cd_3As_2$\cite{Neupane2014,Liu2014_2} semimetals. Since the realization of WSM
requires breaking of IS or TRS or both, it has been predicted theoretically that
pyrochlore iridate materials\cite{Wan2011}, topological multilayer
structures\cite{Wan2011} and TaAs class of
systems\cite{Huang2015,Lee2015,Sun2015} can host Weyl nodes. Shortly after these
predictions, the existence of WSMs was verified in experiments with
TaAs\cite{Lv2015} and TaP\cite{Xu2015}.

The degeneracy associated with WP does not rely on any
particular symmetry other than the translation symmetry of the crystal lattice.
Each WP is characterized by a topological charge of
a definite (positive or negative) chirality, corresponding to a bulk Chern number. 
This makes the properties of the bulk electronic bandstructure topologically robust
against external perturbations. Moreover, as a result of the bulk-edge
correspondence, the non-trivial bulk topology of a WSM
gives rise to robust Fermi arc surface
states\cite{Wan2011,Huang2015,Xu2015,Xu2015_2,XuN2016}.  
Weyl fermions in WSMs are predicted to possess unusual
transport phenomena\cite{Shuo2017}. For example, in the bulk they can
give rise to negative magnetoresistance, anomalous Hall effect, non-local
transport and local non-conservation of ordinary
current\cite{Zyuzin2012,Liu2013,Hosur2013,Parameswaran2014}. Weyl Fermi arc
surface states on the other hand are predicted to show novel quantum oscillations in
magnetotransport and quantum interference effects in tunnelling
spectroscopy\cite{Hoser2012,Ojanen2013,Potter2014}. 
Similar to WSMs, double
Fermi arc surface states are also observed in DSMs\cite{Xu2015_3} but they may
not have any topological protection in general\cite{Mehdi2016,Wu2019}.

Apart from naturally occurring WSM materials, such as the
TaAs class, it is interesting and important to investigate materials that can
become WSMs as a result of topological phase transitions induced by external
perturbations.  For example, phase transitions from DSMs to other topological
phases, included WSMs, were discussed by Yang {\it et al.,} using theoretical
models\cite{Yang2014}. The simplest way to turn a DSM into a WSM is to apply an
external magnetic field, which breaks TRS. In $Cd_3As_2$, a magnetic
field-driven splitting of Landau levels and a nontrivial Berry phase were
detected\cite{Cao2015}, which are consistent with the Weyl phase.

On the other hand, considerably less investigated are the topological phase
transitions induced by doping DSMs. Different types of impurities and the ways
in which they are incorporated in pristine DSM materials, can selectively break
the symmetries that are required for the stability of the Dirac nodes in a DSM.
This leads to a variety of phase transitions to both topologically trivial or
non-trivial phases. First-principles studies\cite{Narayan2014} of $Na_3Bi$ and $Cd_3As_2$ DSMs
alloyed with Sb and P, carried out within coherent potential
approximation (CPA) where the crystal symmetries are preserved, show that
these materials remain in the DSM phase up to 50\% concentration before making
a transition to a trivial insulator\cite{Narayan2014}. More recently, the effect of
magnetic impurities in  DSMs has been studied theoretically within a model
Hamiltonian approach, showing that the breaking of TRS by magnetic impurity
potential splits a Dirac node into two Weyl nodes\cite{Ming2017}. 

In this work, using DFT methods, we have systematically investigated the effect
of nonmagnetic zinc (Zn) and magnetic manganese (Mn) impurities in the
prototypical $Cd_3As_2$ and $Na_3Bi$ DSMs.
In particular, we have carefully introduced impurities
to break selectively different symmetries, and have analyzed the consequences of
terms that break individual or multiple symmetries on the topological
properties. We have also investigated the combined effect of doping and strain
on the topological properties. Our work shows that a DSM makes transitions to a
WSM, a topological insulator or an ordinary insulator phase, depending on which
symmetry is broken. 
Importantly, we find that in doped
$Cd_3As_2$ where IS is broken by nonmagnetic impurities while TRS and rotational
symmetry  are preserved, modified DSM and WSM phases can even coexist at special
points of the parameter space, which can be reached by applying an external
strain. This occurrence bears some similarities to the mixed phase recently
found theoretically in polar hexagonal $ABC$ Crystal SrHgPb\cite{Gao2018}. When
TRS or both TRS and IS are broken, the system makes a transition to a magnetic
Weyl phase. 

The paper is organized as follows. In Sec.~\ref{computation} we
describe the details of the computational approach, which include the DFT
methods and the topological analysis based on the calculations of different
topological invariants carried in atomistic tight-binding models  extracted form
DFT. In Sec.~\ref{result}  we present the results for pure $Cd_3As_2$ and the
consequences of different symmetry breaking, and in Sec.~\ref{NaBi_DSM} we have
discussed our calculations of antimony (Sb) doped $Na_3Bi$ DSM. Finally, in
Sec.\ref{conclusions} we present the conclusions and outlook.

\section{Computational details} 
\label{computation}

We have used three different computational tools to study the electronic and the
topological properties of $Cd_3As_2$ and and $Na_3Bi$ DSMs. \\

{\it DFT --} To perform electronic structure calculations, we have first relaxed
the crystal structure for both the cell parameters and the atomic positions
using the Quantum Espresso ab-initio code\cite{QE}. 
The final relaxed structure is then used to study electronic properties
in the presence of spin-orbit coupling, by employing the full-potential
all-electron linearized augmented plane-waves method as implemented in WIEN2K
ab-initio code\cite{Wien2k}. The Perdew-Burke-Ernzerhof generalized gradient
approximation (PBE-GGA) is used for the exchange correlation
functional\cite{Perdew1996}. For a few calculations (Mn doping) we have employed 
the Vienna Ab Initio Simulation Package (VASP)\cite{vasp1,vasp2}.
We have checked a few benchmark cases with both Wien2k and VASP, 
finding that the two DFT codes give consistent results. 

The crystal structure of $Cd_3As_2$ at ambient conditions has the tetragonal
symmetry D$^{15}_{4h}$ (P$_{4_2}/nmc$), with a 40-atom unit cell\cite{Atom40}.
The symmetries that play a crucial role in the electronic properties of this
crystal are TRS, IS, two mirror planes (M$_{xz}$ and M$_{yz}$), and a
nonsymmorphic screw symmetry S$_{4z}$ (consisting of a 4-fold rotation about the
z-axis followed by a half lattice translation along the z-axis)). We will see
later that this screw axis is essential for the stability of the DPs. Note that
apart from these symmetries, the point group of the crystal also contains
dihedral mirror planes and $C_2$ rotation axes.

We have constructed the cell using the experimental lattice constants and have
relaxed both the cell parameters and the atomic positions until the forces are
less than 1 mRy/au and  the stress on the cell is less than 0.5 Kbar. The energy 
convergence was set to $10^{-3}$ mRy.
This calculation is performed using Quantum Espresso with the
cutoff energies E$_{\rm wfc}$ = 80 Ry for the wave function and E$_{\rho}$ = 600
Ry for the charge density and potential; a uniform Monkhorst-Pack mesh of 7x7x5
k-points has been used. The relaxation increases the cell parameters from
($a$=$b$=16.89, $c$=23.96 bohr) to ($a$=$b$=17.27, $c$=24.27
bohr). The atomic positions also change after relaxation, but not very
significantly. This relaxed structure is the basis of the remaining calculations performed in this
work. 

The topological phase transitions are studied by introducing different
realizations of substitutional Zn or Mn impurities at Cd sites in order to break
different symmetries of the system. We have chosen Zn and Mn as substitutional impurities for Cd because
these elements have the same valence states of Cd (two electrons in the s states) and
atomic radii close to the one of Cd, which minimizes the stress induced by
alloying. Note that Zn and Mn are usually substitutional impurities of Cd in II-VI
compounds, and both alloys, Cd$_{3-x}$Zn$_x$As$_2$ and Cd$_{3-x}$Mn$_x$As$_2$ have
been  synthesized \cite{nishihaya2018,Mekyia2019,sun2019}. 

Before analyzing different cases, a few comments on the 
how impurity doping is introduced in the crystal in the first-principles calculations
are in order. Although for a given impurity concentration several structural
configurations are in principle possible, the constraint of breaking one particular symmetry while
preserving the others reduces significantly the number of different
realizations that are actually allowed. Even when we have more than one
disordered realization for breaking a given symmetry, those
configurations themselves are related by the remaining symmetries of the crystal,
and therefore they are not expected to provide significantly different results. 

However, in the case where both inversion and rotational symmetries are broken by substituting 
only one Cd by a Zn atom (this will be the case 
leading to a Weyl phase), there exist multiple, genuinely-different, doping configurations that
should be addressed. Since these calculations are very time consuming, we have limited ourselves to consider
only two different disordered configurations. 
Typically we find that, except for a small difference in the bandstructure, 
the topological properties are the same for both configurations.

\begin{figure}[h]
\centering{\includegraphics[width=0.5\textwidth]{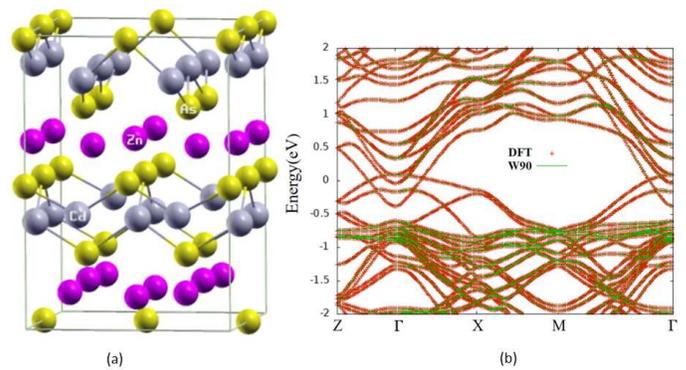}} 
\caption{a)
Relaxed structure of $Cd_3As_2$ doped with 12 Zn atoms to break IS while
preserving S$_{4z}$. This type of doping is known as $\delta$-doping and is
easily realizable experimentally. b) Comparison of the Zn-doped $Cd_3As_2$
bandstructure obtained from DFT and that from the real-space tight-binding
Hamiltonian in the maximally localized Wannier basis. The good agreement between
these two methods demonstrates that the Wannierization procedure is
satisfactory. } \label{CdAs_struct} \end{figure}

The different types of chemical doping and the different impurity concentrations
that we have considered in this work are the following:

\begin{enumerate} \item 4 Zn atoms in the 40-atoms unit cell, with 24 Cd atoms
($\sim $17\% of dopants), placed in order to maintain inversion symmetry (IS)
while breaking the S$_{4z}$. We have also considered 2 Zn atoms ($\sim $9\% of
dopants) to preserve IS and $C_2$ while breaking the S$_{4z}$. 
  
\item  1 Zn atom ($\sim$ 4\% of dopants), in order to break both IS and
S$_{4z}$. 

\item 12 Zn atoms (50\% of dopants), which simulates $\delta$ doping (dopants
placed on planes perpendicular to the rotation axis), in order to break IS while
preserving S$_{4z}$ (see Fig~\ref{CdAs_struct}a).  

\item 2 Mn atom ($\sim$ 8\% doping), in order to break TRS and 1Mn atom to break
both IS and TRS.  \end{enumerate}

After relaxing the system for each of these impurity cases, we have investigated
their projected band structure using the Wien2k code.

{\it Wannier90 --} For the topological studies of this system, we have
constructed a real-space tight-binding (TB) Hamiltonian in the basis of the
Wannier states. The Wannier90 code\cite{Wannier90} is used to construct 
Maximally Localized Wannier Functions (MLWFs) from the Kohn-Sham
orbitals obtained from the Wien2k calculations. Since the DFT calculations show
that the states near the Fermi level are predominantly 5$s$ of Cd, 4$s$ and 4$p$
of As, and 4$s$ of Zn atoms (more detailed discussion is in
section~\ref{result}), we initially projected the Bloch states on these
orbitals. Furthermore, taking into account the spin-orbit coupling, we have
considered the Wannier functions as spinors, namely two component states, and
therefore a total of 176 Wannier functions have been chosen. In the case of Mn
magnetic impurities, we have taken into account also their 3$d$ orbitals, which
increased the total Wannier functions up to 186.    

The accuracy of the calculation of the topological properties relies on the
accuracy of MLWFs. We have used two criteria for acceptable accuracy: (i) the
spread of the Wannier functions should be smaller than the smallest lattice
constant; (ii) the bands calculated  from the Wannier Hamiltonian should be a
good match with the DFT bands. Furthermore, to reduce the numerical error during
the Wannierization, we have also implemented disentanglement, a procedure to
project out the contribution of the relevant from unwanted
bands\cite{Souza2001}. Fig~\ref{CdAs_struct}b shows the DFT and Wannier90 bands
for the impurity case 3 mentioned above. A good match reflects the fact that
satisfactory MLWFs have been achieved. 

{\it WannierTools --} The real space Hamiltonian obtained from Wannier90 code is
then used in WannierTools\cite{WTools} to study different topological properties
of the system. The code is used to search for the nodes in the Brillouin zone
(BZ), and to calculate different topological indices such as $Z_2$, Chern number
etc., which helps to identify different topological phases. In WannierTools, the
$Z_2$ and the Chern number are calculated using the Wannier Charge Center (WCC)
method\cite{Soluyanov2011}. According to this method, a hybrid Wannier function
is constructed for each band by integrating out one component of $\bf k$ vector, say
$k_z$. The WCC of band $n$ is then the expectation value of $r_z(k_x, k_y)$ in
this hybrid Wannier function. The evolution of the WCCs along a k-path in a
given plane, say the $k_x-k_y$ plane, of the BZ can be used to calculate the
topological properties of the plane. For details see
Ref.~\onlinecite{Soluyanov2011}.

A similar strategy has been also employed to study the electronic and
topological properties of $Na_3Bi$. The unit cell of hexagonal $Na_3Bi$ crystal
contains eight atoms (two Bi and six Na atoms). It turns out that the crystal
structure is such that in order to break one specific symmetry without breaking
the remaining ones (for example, breaking the screw symmetry while maintaining
the IS) requires considerably large supercells. A large supercell requires many
Wannier functions for achieving an acceptable Wannierization procedure, and it
also introduces many bands in the first BZ due to band folding, which makes it
difficult to study the topological properties reliably. For this reason, we have
considered only the case with broken IS.

\section{Topological phase transitions in doped $Cd_3As_2$ DSM} 
\label{result} 

The presence of IS and TRS in pure $Cd_3As_2$
leads to the double degeneracy of each band. When band inversion between the
conduction band (primarily consisting of Cd $s$ states) and the valence band
(primarily consisting of As $p$ states) occurs, as shown in Fig~\ref{NoC4band}a,
two Dirac nodes appear symmetrically around the $\Gamma$-point on the Z-$\Gamma$
axis at (0, 0, $\pm$ 0.081)(1/\AA), consistent with previous
works\cite{Borisenko2014,Zhou2016,Mosca2017,Crasse2018}. The two DPs lie at
-0.008 eV, slightly below the Fermi energy. The system can be further
characterized by the $Z_2$ invariant of the six time-reversal-invariant
planes (TRIPs). Since the Dirac nodes are on the $k_z$ axis, only $k_z$=0 and $k_x, k_y, k_z$
= $\pi$ planes are gapped. Our calculations show that $Z_2$=1 only for the $k_z = 0$-plane,
implying that the associated gap is non-trivial. We have further verified the
existence of {\it double Fermi arcs} surface states, as shown in the inset of
Fig~\ref{NoC4band}a. These results serve as a benchmark for the calculations of
the doped cases described below.

\subsection{Non-magnetic impurities in $Cd_3As_2$}

In this section we study the effect of doping 
on the electronic and topological properties of $Cd_3As_2$, when
IS, $C_2$ rotational symmetries, and $S_{4z}$ screw
symmetry, are selectively broken, while TRS is preserved. 
This is achieved by introducing substitutional
non-magnetic Zn impurities at the Cd sites. Here we discuss three different impurity
realizations and their consequences on the topological properties.\\  

\subsubsection{Broken $S_{4z}$ screw symmetry: transition to $Z_2$ semimetal and insulator phases}
\label{Z2phase}

\begin{figure}[h]
\centering{\includegraphics[width=0.48\textwidth]{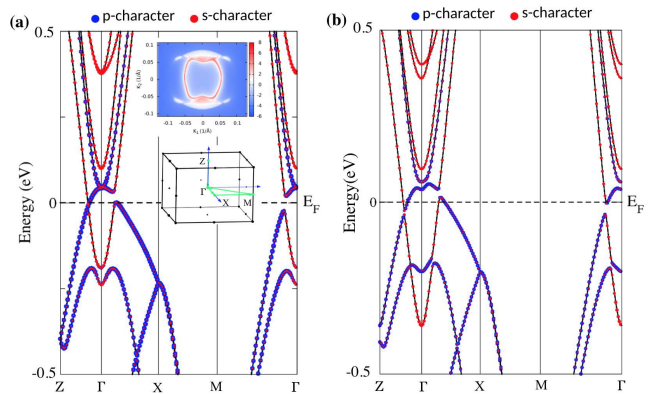}} \caption{(a) The
projected bandstructure of $Cd_3As_2$. The presence of Cd $s$ states (red) below the
Fermi level and As $p$ states (blue) above the Fermi level
indicates that bands are inverted around the $\Gamma$-point, leading to
nontrivial topology. The inset shows the existence of {\it double Fermi arcs}
surface states. (b) The projected bandstructure of Zn doped $Cd_3As_2$ with
broken $S_{4z}$ screw symmetry. Band inversion is still preserved as in
case of (a), signaling a non-trivial topology, but a gap opens up at the DP.}
\label{NoC4band} \end{figure}

In order to break the $S_{4z}$ screw symmetry while preserving IS (and TRS), we have
substituted four Cd atoms by 4 Zn atoms. Since both TRS and IS are still
present, all bands are two-fold degenerate. However, by breaking the screw
symmetry we also break other symmetries such as mirror planes and, consequently,
the space group reduces to P-1, which contains only the IS. At a general point
on the $k_z$ axis, the point group is $C_1$, which contains only the unit
operator. Therefore, both the conduction and valence band states belong to the $A_1$
irreducible representation of this point group, and they are allowed to mix when
they approach each other. Consequently, the Z-$\Gamma$ path becomes gapped, as shown
in Fig~\ref{NoC4band}b. Our DFT calculations show that the smallest gap is of
the order of 40 meV at around (0, 0, $\pm$ 0.095)(1/\AA), which is slightly away from
the DPs of the pure $Cd_3As_2$.    

The projected orbital states in Fig~\ref{NoC4band}b shows that the valence band
is predominantly of As $p$ character (blue) whereas the
conduction band is predominantly of Cd $s$ and Zn $s$ character (red). It is also
evident that around the $\Gamma$-point bands are inverted as there is a
considerable presence of $s$ states just below the Fermi level and $p$ states
above the Fermi level, which gives rise to the non-trivial topology. 

We note from the bandstructure that this system is not a proper insulator, but
rather a semimetal, since the Fermi level crosses both the conduction and
valence bands, creating an electron pocket on Z-$\Gamma$ path and a hole pocket
on $\Gamma$-X path, respectively. However, since the valence band has a finite
{\it direct} bandgap everywhere in the BZ, it can be topologically classified in
the same way as it is done for bismuth and antimony semimetals\cite{Hasan2010}.
Therefore, we have calculated the $Z_2$ indices for all the six TRIPs
($k_x=0,\pi; k_y=0,\pi; k_z=0,\pi$). We find that $Z_2$=1 for the
$k_x, k_y, k_z=0$ planes and zero for all other planes. The topological index of
the system is then $\nu$=(1;0,0,0), which is the same index of a strong 3D
topological insulator. Although this system is considered as a semimetal due to
the presence of a few states at the Fermi level, the non-trivial $Z_2$ topology
suggests the existence of non-trivial surface states for this system.   

To further confirm that $S_{4z}$ is necessary for the stability of the Dirac
nodes, we have performed an additional  calculation where both IS and $C_2$ are
preserved but $S_{4z}$ is broken (space group $P2_1/m$), which is realized by
substituting two Cd by two Zn atoms. In this case, the Fermi level is fully
gapped everywhere in the BZ, with the smallest gap being of the order of 30 meV at the
original location of the DPs. The bandstructure is similar to that of
Fig~\ref{NoC4band}b, except that the bands do not cross the Fermi level. The
topological analysis shows that this system is also a strong TI with
$\nu$=(1;0,0,0).

These calculations clearly demonstrate the importance of the rotational $S_{4z}$ screw
symmetry for the stability of the DPs in the $Cd_3As_2$, consistent with the
stability criteria discussed in \cite{Yang2014}. They also show how breaking this
symmetry can cause a transition to a topologically distinct phase.

\subsubsection{Broken IS and $S_{4z}$ symmetry: transition to a Weyl phase} 
\label{weyl_phase}

\begin{figure}[h]
\centering{\includegraphics[width=0.46\textwidth]{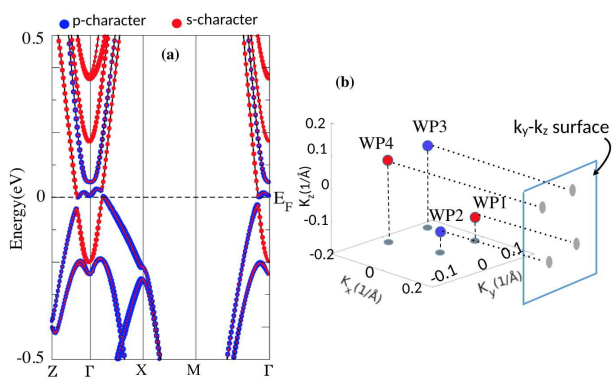}} \caption{(a)
The bandstructure of $Cd_3As_2$ with broken IS and $S_{4z}$ screw symmetry. 
As $p$ and Cd $s$ states are highlighted with blue and red, respectively. (b) Each
Dirac node splits into two Weyl nodes of opposite chirality. The red (blue) dots
represent Weyl nodes of +1 (-1) chirality. The projection of the Weyl nodes on the $k_x-k_y$
and $k_x-k_y$ planes is also shown.} 
\label{NoC4NoIband} 
\end{figure}

In this section we discuss the effect of breaking both IS and $S_{4z}$ screw
symmetry. This can be achieved simply by replacing one Cd atom by one Zn atom in
the unit cell. As a result of this impurity configuration, the  space group of
the crystal changes to Pm space, which contains only a reflection plane. Since
the IS is absent in this space group, the bands are no longer doubly degenerate,
except at the TRIMs. Consequently,  Dirac nodes cannot form at any high symmetry
line. Indeed, a gap of 30 meV opens up at the location of the DP of the pure
system, as shown in Fig~\ref{NoC4NoIband}a. However, this opens up the
possibility that each Dirac node may split into two Weyl nodes.
To search for the Weyl nodes, we have constructed the real-space TB Hamiltonian
in the WF basis as described in  section~\ref{computation}. We have then
searched for the nodes between the highest occupied and the lowest unoccupied
bands in the entire BZ using WannierTools. We found four nodes in the BZ at
$k$-values (0.067, $\pm$ 0.025, -0.042)(1/\AA) and (-0.067, $\pm$ 0.025, 0.042)(1/\AA)  
and energy E=-0.003 eV, as shown in Fig~\ref{NoC4NoIband}b. To characterize the topological
nature of these nodes, we have calculated their chirality. We find that two of
the nodes have chirality +1 (red) and the other two have chirality -1 (blue),
which we tentatively interpret as 4 Weyl nodes originating from the two original
Dirac nodes. Note that the nodes are closer to the $\Gamma$ point compared to
the location of the Dirac nodes in the pure $Cd_3As_2$.  

\begin{figure}[h]
\centering{\includegraphics[width=0.46\textwidth]{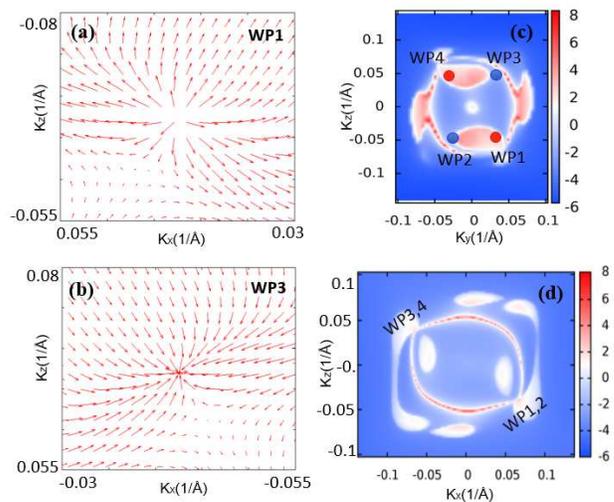}} 
\caption{
The Berry curvature in the $k_x-k_z$ plane, at fixed $k_y = 0.025$(1/\AA)
for the doping case considered in Fig.~\ref{NoC4NoIband}. 
(a) and (b) The Berry curvature in the neighborhood of WP1 and WP3 of 
Fig.~\ref{NoC4NoIband} with opposite chiralities, respectively. the curvature diverges
at the location of the WPs. (c) Fermi arc surface states projected onto the (100) 
surface. (d) The Fermi arc states projected onto (010) surface.} 
\label{NoC4weyl} 
\end{figure}

To further confirm that these nodes are indeed Weyl points, we have calculated
the Berry curvature, since a Weyl point acts as a source or a drain of Berry
curvature in momentum space. In Fig.~\ref{NoC4weyl}(a,b) we have plotted the
curvature in the $k_x-k_z$ plane, at fixed $k_y = 0.025$(1/\AA), around the
two Weyl nodes WP1 and WP3 ( Fig~\ref{NoC4NoIband}b), respectively. It is evident 
that WP1 and WP3 act as a source and drain, respectively, as the Berry curvature 
diverges at the WPs, which supports the conclusion that these are two Weyl nodes 
of opposite chirality. 
In Fig.~\ref{NoC4weyl}(c), we have plotted the surface states spectra on
the (100) surface at constant E=-0.003 eV.  We can clearly see
two Fermi arcs, emanating from two different Weyl points, although some bulk states 
are also present. We have also plotted surface states on (010) surface in 
Fig.~\ref{NoC4weyl}(d). The WPs (WP1, Wp2) and (WP3, WP4) are projected onto the 
the same points on this surface, therefore, the two arc states appear to emanate from
the same point.  

The bands are fully gapped on the $k_z=0$ TRIP, and $Z_2$ = 1 on this plane, which indicates a
non-trivial gap. This opens up the possibility of observing a quantum spin
Hall effect on this plane, similar to that one predicted for the TaAs WSM
family\cite{Sun2016}. Our calculations clearly demonstrate that the insertion
of a single Zn impurity in the $Cd_3As_2$ unit cell causes a topological phase
transition to a Weyl phase, which is more robust than the DSM phase.

\subsubsection{Broken IS symmetry: coexisting of Dirac and Weyl phases} \label{DWSM}

We finally discuss the last example of nonmagnetic doping in $Cd_3As_2$, consisting
of a $\delta$-type doping of Zn impurities (see Fig.~\ref{CdAs_struct}) such
that IS is broken while the $S_{4z}$ screw symmetry is preserved. The dopants
substitute twelve Cd atoms in two different parallel planes perpendicular to the
tetragonal axis (along the z direction), resulting in $I4_2mc$ space group
(associated point group, $C_{4v}$). This space group contains the $M_{xz}$ and
$M_{yz}$ mirror planes along with the $S_{4z}$ screw axis. 

This is a particularly interesting case. Since $M_{yz}$ and $S_{4z}$ are
symmetries of this space group, they commute with the Hamiltonian:
$[M_{yz},H]=0$, $[S_{4z},H]=0$. However, these two symmetries do not commute
with each other, but rather they {\it anticommute},
$\left\{M_{yz},S_{4z}\right\}=0$. The anti-commutation of two symmetry operators
of a Hamiltonian plays a crucial role in generating degeneracies of the energy bands. To
elaborate, let $\psi_s$ is an eigenstate of $S$ with eigenvalue $s$, which is also an
eigenstate of $H$ with energy $E$ i.e. $S\psi_s=s\psi_s$ and $H\psi_s=E\psi_s$.
Then, $H(M\psi_s)=M(H\psi_s)=E(M\psi_s)$. Therefore, $M\psi_s$ is also an
eigenstate of $H$ with energy $E$. Now, since $S$ and $M$ anticommute,
$S(M\psi_s)=-M(S\psi_s)=-s(M\psi_s)$. Therefore, $\psi_s$ and M$\psi_s$ are two
orthogonal eigenstates of $H$ with the same energy $E$, that is, $E$ is degenerate. The
presence of these two anti-commuting symmetry operators ensures that all bands
are two-fold degenerate along the tetragonal axis, $k_z$ (Z-$\Gamma$-Z path
where band inversion occurs in the pure $Cd_3As_2$). Because of the maintained
$S_{4z}$ screw symmetry, the Dirac points of the DSM phase are still present on
this axis in this doped system.

Furthermore, the breaking of IS gives rise to the  possibility of two-fold
degenerate nodes, possibly Weyl points, away from the tetragonal axis. These
nodes cannot be directly generated from the original Dirac points, as in the
case of a single Zn doping, because in this case the original Dirac nodes are
still present in the system. Therefore, this particular implementation of the
impurities, opens up the possibility of observing a Dirac+Weyl coexisting
phase, similar to the one recently discovered in polar hexagonal $ABC$ Crystal
SrHgPb\cite{Gao2018}.     

\begin{figure}[h]
\centering{\includegraphics[width=0.32\textwidth]{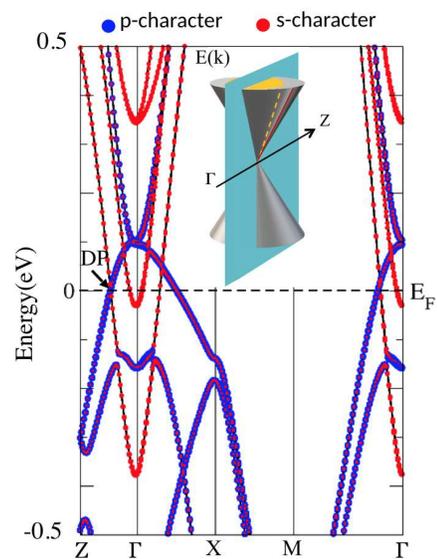}} \caption{The
bandstructure of Zn-doped $Cd_3As_2$ with broken IS but with 
preserved $S_{4z}$ screw symmetry. The projections of As $p$ and Cd $s$ states
are shown in blue and red, respectively. The black arrow indicates 
the DP on the Z-$\Gamma$ path along which the double degeneracy of the bands is 
preserved. The inset shows a
schematic of the bands around the DP, plotted in a $k$-plane containing the
Z-$\Gamma$ axis. The doubly degenerate states of the cone are
those that lie on the shaded 2D plane.} \label{NoIwC4bands} \end{figure}

\begin{figure}[h] \centering{\includegraphics[width=0.46
\textwidth]{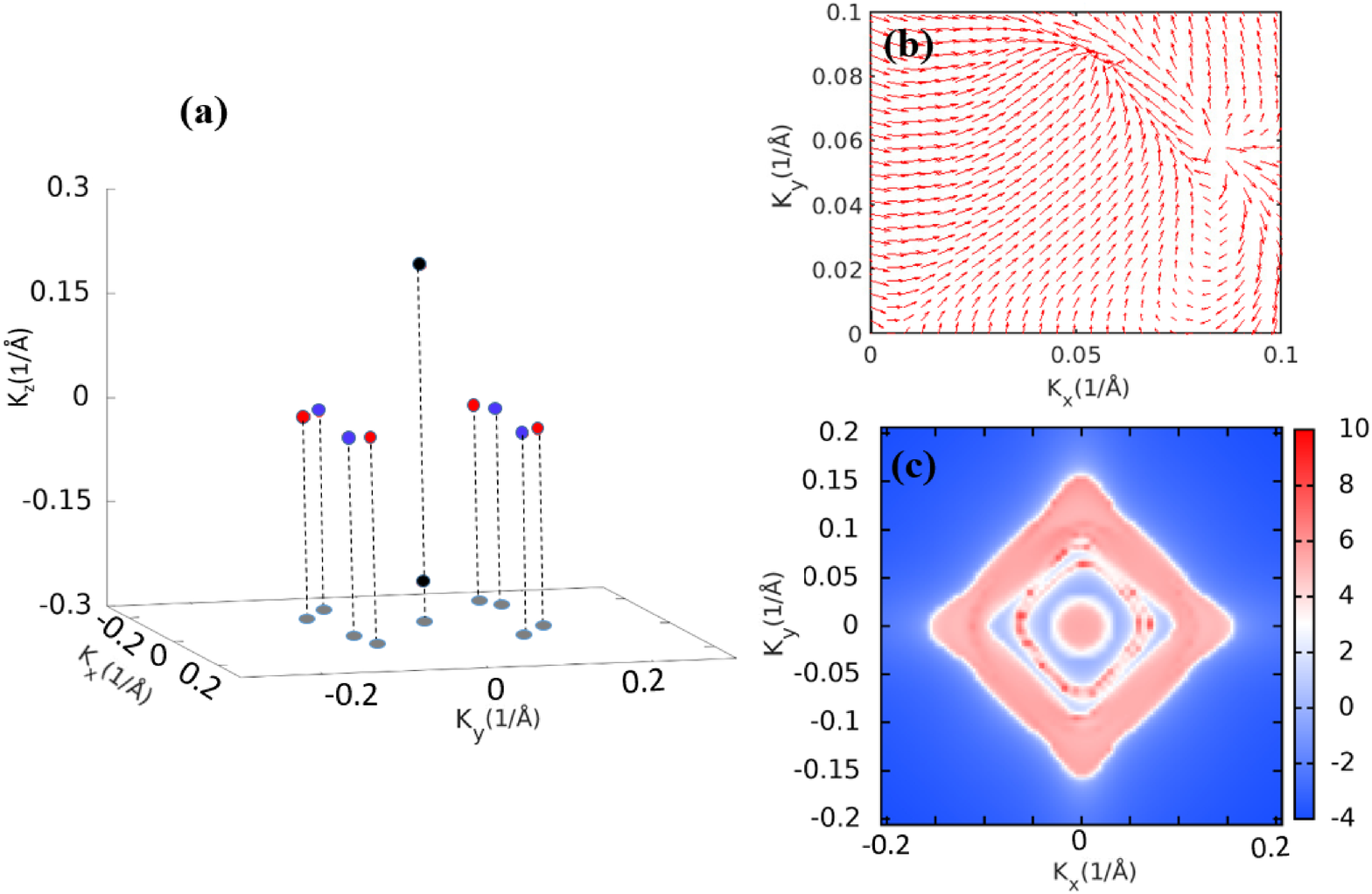}} \caption{(a) Dirac+Weyl coexisting phase: a pair of
Dirac nodes on the $k_z$-axis and four pairs of Weyl nodes on the $k_z =0$ plane (blue
and red dots represent Weyl nodes of opposite chirality). 
(b) The Berry curvature around a
pair of Weyl nodes of opposite chirality, showing the divergence of the curvature
at the nodes. (c) The Fermi are surface states projected onto the (001) surface at
E=0.016 eV above the Fermi level.} \label{Weylpts} \end{figure}

For a detailed analysis, we have performed calculations for both relaxed and
unrelaxed structures. Fig~\ref{NoIwC4bands} shows the bandstructure for the
unrelaxed structure. Here by {\it unrelaxed} we mean the system where the
impurities have been inserted in the {\it relaxed} pure $Cd_3As_2$ crystal,
without carrying out any further relaxation. We can clearly see the presence of
a four-fold Dirac node with zero chirality (derived from WannierTools) along
Z-$\Gamma$-Z path for which the double degeneracy is preserved. This is similar
to the pure $Cd_3As_2$ case, however, it differs in an important way. In pure
$Cd_3As_2$, the linearly dispersed states emanating from the Dirac node form a
doubly-degenerate 3D Dirac cone. But in this case, it can only form a
doubly-degenerate cone on a 2D  plane (the shaded plane shown in the inset of
Fig.~\ref{NoIwC4bands}) for the states along the $k_z$-axis. Away from
Z-$\Gamma$-Z path the bands are no longer doubly degenerate and split into two
nondegenerate cones (the gray cone, and the yellow cone inside it as shown in
the inset).    
To search for the nodes, we need to look for band-touching points at generic
points in the full BZ, not necessarily on the high-symmetry axis, which is difficult
to achieve using DFT. To facilitate this search, we have constructed a TB model in
the Wannier function basis using the Wannier90 code, and then we have used WannierTools
to search for nodes. We found two band touchings along the tetragonal axis
at ${\bf k} = (0,0,\pm 0.11$)(1/\AA), 
namely the expected Dirac points, and eight new two-fold
degenerate nodes in the $k_z = 0$ plane, as shown in Fig~\ref{Weylpts}a. The
Dirac nodes are essentially at the Fermi energy, but the Weyl nodes lie about 16
meV above the Fermi level. To investigate whether these eight nodes are Weyl
nodes, we have calculated the chirality of each node, which shows that four of
the nodes have chirality $+1$ (red dots) and the remaining four have chirality
$-1$ (blue dots), suggesting that they are indeed Weyl nodes. The Dirac nodes,
on the other hand, have chirality zero. For further evidence, we have plotted
the Berry curvature on the $k_z=0$ plane, zoomed around two nodes of opposite
chirality, as shown in Fig~\ref{Weylpts}b. It clearly shows source-like and
drain-like divergences of the curvature around each node, which is a
characteristic feature of the Weyl nodes. In Fig~\ref{Weylpts}c we have plotted
the surface states projected onto (001) plane at E=0.016 eV. While small bulk
states are still present at this energy, it is also evident there are surface
states connecting different pairs of Weyl nodes.  
These results indicate that this unrelaxed $(Cd_{1-x}Zn_x)_3As_2$ (x=0.5) system
is a very unusual and interesting topological system in which both Dirac and
Weyl nodes coexist, resulting in a new topological phase that is a mixture of
DSM-like  and WSM phases (We say, DSM-like because, as we discussed above,  the
bands crossing at these nodes are doubly-degenerate only along the $Z-\Gamma-Z$
path).

In order to further elucidate this mixed phase, we formulate a low-energy
effective Hamiltonian $H(\bf{k})$ using a four-band $\bf k\cdot p$ model. We consider the
conduction and valence bands around the $\Gamma$ point where Cd-$5s$ and As-$4p$
states are inverted. In presence of spin-orbit coupling, these states at the $\Gamma$ point can
be specified by the eigenvalues of the total angular momentum 
operators ${\bf J}^2$ and $J_z$: $\ket{S_{1/2},\pm 1/2}$,
heavy hole states $\ket{P_{3/2},\pm 3/2}$, light hole states $\ket{P_{3/2},\pm
1/2}$ and split-off states $\ket{P_{1/2},\pm 1/2}$. The light hole and split-off
states lie further below the Fermi level; therefore, in order to construct the minimal
four-band model we have considered only the $\ket{S_{3/2},+3/2}$, $\ket{S_{1/2},+1/2}$
$\ket{S_{1/2},-1/2}$, and $\ket{P_{3/2},-3/2}$ states.  

Applying $\bf k\cdot p$ perturbation theory to these
states up to the second order, and making use of the crystal symmetries,
the Hamiltonian  $H(\bf{k})$ takes the following form \cite{Wang_2013}
%
\begin{equation} H(\boldsymbol{k})=\epsilon_0(\boldsymbol{k})+\begin{pmatrix}
-M(\boldsymbol{k}) & Ak_- & 0    & 0     \\ 
Ak_+ & M(\boldsymbol{k})  & Dk_- & 0     \\ 
  0  & Dk_+ & M(\boldsymbol{k})  & -Ak_- \\ 
  0  & 0    & -Ak_+ & -M(\boldsymbol{k})
\end{pmatrix} 
\label{Ham} 
\end{equation}
where $\epsilon_0(\boldsymbol{k})= C_0 + C_1k_z^2 + C_2(k_x^2+k_y^2)$,
$M(\boldsymbol{k})=M_0 - M_1k_z^2 - M_2(k_x^2+k_y^2)$ and $k_\pm=k_x \pm ik_y$. The parameter
D reflects the breaking of IS. If $D=0$, Eq.~\ref{Ham} can be linearized about the DP to obtain 
the Dirac Hamiltonian of pure $Cd_3As_2$. The energy dispersions are
\begin{equation} \begin{split} E_1^{(\pm)}(\boldsymbol{k}) =
\epsilon_0(\boldsymbol{k}) + \frac{1}{2}\biggl[\pm D\sqrt{k_x^2+k_y^2} & \pm  \\
\sqrt{(4A^2+D^2)(k_x^2+k_y^2) \pm
4M(\boldsymbol{k})D\sqrt{k_x^2+k_y^2}+4M(\boldsymbol{k})^2)}\biggr] & \\
E_2^{(\pm)}(\boldsymbol{k}) = \epsilon_0(\boldsymbol{k}) + \frac{1}{2}\biggl[\pm
D\sqrt{k_x^2+k_y^2} & \mp  \\
\sqrt{(4A^2+D^2)(k_x^2+k_y^2)\pm4M(\boldsymbol{k})D\sqrt{k_x^2+k_y^2}+4M(\boldsymbol{k})^2)}\biggr]
& \\
\end{split} 
\label{disp1} 
\end{equation}

\begin{figure}[h]
\centering{\includegraphics[width=0.48\textwidth]{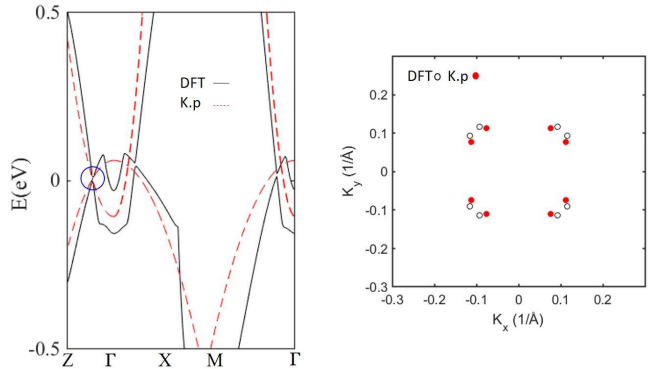}} \caption{(a)
Comparison of the DFT band structure and the bandstructure given by the ${\bf k}\cdot {\bf p}$ model 
for the coexisting phase.
The blue circle indicates the Dirac point. (b) Position of the Weyl nodes obtained from DFT (open circles)
and the ${\bf k}\cdot {\bf p}$  model (filled red circles).} \label{kp} \end{figure}

The coefficients in Eq.~\ref{Ham} are obtained by fitting to the DFT bands and are listed in
Table~\ref{Coeff}. Fig.~\ref{kp}a compares the bands from DFT and
${\bf k}\cdot {\bf p}$ model, which are in a good agreement. Along the Z-$\Gamma$-Z
axis (0, 0, $k_z$) these dispersion energies are two-fold degenerate, and when
they cross, they form a Dirac node. The positions of the Dirac nodes obtained
from Eq.~\ref{disp1} are (0, 0, $\pm \sqrt{\frac{M_0}{M_1}}$). To check the topological natural of 
these nodes, in particular if they have a chiral charge, we perform a Taylor expansion
of the dispersion around the Dirac point. 
Keeping only terms linear in the variation $\delta {\bf k}$, we obtain
\begin{equation} \begin{split}
E_1^{(\pm)}(\delta\boldsymbol{k})=\pm \abs{v_{1 xy} (\delta k_x\hat{\boldsymbol{x}} + \delta k_y\hat{\boldsymbol{y}}) + 
v_{z}\delta k_z\hat{\boldsymbol{z}}} \\
E_2^{(\pm)}(\delta\boldsymbol{k})= \mp \abs{v_{2 xy}(\delta k_x\hat{\boldsymbol{x}}+ \delta k_y\hat{\boldsymbol{y}}) +
v_{z}\delta k_z\hat{\boldsymbol{z}}}
\end{split} 
\label{disp2} 
\end{equation}
where the two in-plane velocities are $v_{1 xy} =  \frac{1}{2}(+D +
\sqrt{4A^2+D^2})$, $v_{2 xy} =\frac{1}{2}(-D + \sqrt{4A^2+D^2})$, with $v_{1 xy} > v_{2 xy} > 0$, and
$v_{z}= 2\sqrt{{M_0}{M_1}}$.   
The two energy pairs $E_1^{(\pm)}$ and
$E_2^{(\pm)}$ in Eq.~\ref{disp2} are the solutions of the two $2\times 2$ Weyl Hamiltonians 
$H_{WP, 1}(\delta \boldsymbol{k})= {\bf V}_1(\delta \boldsymbol{k}) \cdot{\boldsymbol \sigma}$,
and $H_{WP, 2}(\delta \boldsymbol{k})= -{\bf V}_2(\delta \boldsymbol{k}) \cdot{\boldsymbol \sigma}$ respectively,
where 
${\bf V_{i}}(\delta \boldsymbol{k}) =  
v_{i xy}(\delta k_x\sigma_x + \delta k_y\sigma_y) + v_z\delta k_z\sigma_z$,  $i =1, 2$
and $\boldsymbol{\sigma}$'s are Pauli matrices. 
The minus sign in front of $H_{WP, 2}$ is important: it implies that for each $\bf k$
the two eigenstates corresponding to a given sign of the dispersion are swapped with respect to $H_{WP, 1}$. 
Then the full $4 \times 4$
Hamiltonian around the Dirac point can be written in the form:
\begin{equation}
H_{DP}(\delta \boldsymbol{k})=\big(\begin{smallmatrix} \boldsymbol{V_1}(\delta {\bf k}) \cdot {{\boldsymbol\sigma}}
& 0\\ 0 & -\boldsymbol{V_2}(\delta {\bf k}) \cdot{{\boldsymbol\sigma}} \end{smallmatrix} \big), 
\end{equation}
which corresponds to two coincident Weyl nodes of opposite chirality. Therefore, the Dirac
nodes of this coexisting phase have chiral charge zero. It is also evident from
the Hamiltonian $H_{DP}(\delta \boldsymbol{k})$ that the double degeneracy of
the Dirac cone is preserved only along $k_z$ but splits for $\bf k$-points away from
this tetragonal axis, as obtained from the DFT calculations
(Fig.~\ref{NoIwC4bands}).

In addition to the Dirac nodes, the model, with the {\it same} parameters, 
also predicts the existence of four
pairs of nodes on the $k_x-k_y$ plane.  As shown in Fig~\ref{kp}b, the position
of these nodes essentially coincides with the positions the four pairs of Weyl nodes obtained from the DFT calculations. The small
difference is due to the fact that we are using a simplified  model with just four
bands, which cannot capture all the details of the DFT results. 

\begin{table}[h] \caption{Coefficients for the four-band ${\bf k}\cdot {\bf p}$ model 
of Eq.~\ref{Ham} obtained by fitting the DFT results.}
\centering \vspace{1mm}\scriptsize \begin{tabular}{cccccccc}\hline \hline $C_0$
& $C_1$      & $C_2$      & A        & $M_0$   & $M_1$      & $M_2$      & D
\\ (eV)   & (eV-\AA$^2$) & (eV-\AA$^2$) & (eV-\AA) & (eV)    & (eV-\AA$^2$) &
(eV-\AA$^2$) & (eV-\AA)  \\  \hline		-0.0250 & 28.0234    &145.0648
& -0.0060  & -0.0833 & -82.9471   & -179.7532  & 0.2000   \\  \hline
\end{tabular} \label{Coeff} \end{table}

The effective model can be also used to investigate the stability of the
coexisting phase. We find that this phase is most sensitive to the parameters $A$
and $D$, which are proportional to the in-plane components of the velocity. Our
calculations show that for a small range of $A$ (-0.002 - -0.007) eV-$\AA$ and $D$
(0.06 - 0.265) eV-$\AA$ both Dirac and Weyl nodes remain stable, if all other
parameters are kept fixed to their optimal values listed in Table~\ref{Coeff}. 
We can imagine that changing these parameters might correspond to acting externally on the system,
e.g., by applying an external strain, while maintaining the symmetry of the system.
Although establishing a direct connection between the parameters of the ${\bf k}\cdot {\bf p}$ 
model and external strain is
not straightforward,
these results indicate that at least for a
small range of the parameters that might be controlled externally, the coexisting phase of DSM and WSM should
remain stable.

\subsubsection{Effect of external strain on the coexisting phase}

To further investigate the stability of this mixed phase, we have relaxed the
structure for both the cell parameters and the atomic positions. Relaxation
reduces the cell parameter to $a = b = 16.567031$  and $c = 23.487859$ bohr and
brings the As layer closer to the Zn layer (further away from the Cd layer) but
the bandstructure is  similar to that of unrelaxed case. To search for the nodes
in the full BZ, we have followed the same procedure used above. In this case we
have found only the Dirac nodes along the tetragonal axis (slightly moved along
the line), whereas the Weyl nodes on the $k_z=0$ plane have disappeared. Since
the bands are now gapped on $k_z=0$ plane, we have calculated the $Z_2$
invariant, which shows the plane is topologically trivial. Furthermore, since
$k_z=0$ plane is not a mirror plane of the $C_{4v}$ point group, this plane
cannot be characterized by mirror Chern number either.  Therefore, in this
particular case, the non-trivial topology related to the Dirac phase is broken.
Although the four-fold degenerate nodes are still present in the system
protected by the $S_{4z}$ screw symmetry, no other topological aspects are found
in the system, resulting in a trivial semimetal.  

The relevant question now is whether or not the Dirac+Weyl mixed phase can exist
in more general conditions (as our ${\bf k}\cdot {\bf p}$ model suggests), in particular for
physical configurations where the atomic positions are relaxed, which can be
realized experimentally. In order to realize such a coexisting phase, we need to
consider an additional degrees of freedom, which can induce additional
accidental double degeneracies away from the Z-$\Gamma$-Z axis. 

Since the unit-cell lattice constants $a,\ b,$ and $c$ of the relaxed system are
smaller than the ones of the unrelaxed structure, we have investigated whether
or not the coexisting Dirac+Weyl phase reappears when the unit cell size is
progressively increased from the relaxed one. This procedure is supposed to
describe an applied external strain. Specifically, in these calculations we have
simulated an external strain by acting on the cell in the following way: we first
modify (increase) the cell parameters, and then we relax the atomic positions
for the new cell parameters. We have focused our search by varying the applied
strain in a limited area of the parameter space, namely by increasing the
lattice constants $a=b$ while keeping  the  ratio $c/a$ equal to the value of
the relaxed system $c/a = 1.417747$. The different strained systems that we
considered are listed in Table~\ref{tab_strain}. Note that the first entry in
this table is for the case in which $a$ and $b$ take the value of the unrelaxed
system but the atomic positions have been relaxed. 

We should remark that this particular approach of applying strain may be
difficult to realize experimentally but here we would like like to investigate,
theoretically, whether by progressively increasing the strain close to the
unrelaxed system parameters, the trivial phase of the relaxed structure
eventually changes into a topological Dirac phase. We are interested in how the
band structure changes under strain, looking, in particular, at the gaps around
the Fermi energy which, at some point, must close in order for the topology to
change. A more realistic modeling of strain would certainly be required to 
make predictions that can be verified experimentally.
However, our simplified approach can still provide insight
into the possibility of obtaining and controlling the co-existing phase by external probes.

\begin{table}[h] \caption{Applied strain in $Cd_3As_2$. Strain is modeled by
changing the lattice constants $a$, $b$ and $c$ under the conditions that 
$a=b$, and the ratio $c/a$ is kept fixed and equal to the value of the relaxed
doped structure, 1.42. For the case of the first row (strain I), $a= b = 17.27$ equal
to the value of the unrelaxed system, but the atomic positions have been
relaxed.} \centering \vspace{1mm} \begin{tabular}{|c|c|c|c|c|}
\hline Structure & Lattice constant   & \multicolumn{2}{c|}{Stress (Kbar)}  &
phase     \\ \cline{2-3} &  a = b (bohr)      & along x, y     &  along z   &
\\ \hline I   &  17.27         & -44.90         & -43.43     &  Trivial  \\
II   &  17.30         & -45.00         & -42.56     &  DSM      \\ III   &
17.34         & -47.09         & -44.47     &  DSM      \\ IV    &
17.45         & -51.19         & -47.97     &  DSM      \\ V    &  17.62
& -55.36         & -50.23     &  DSM      \\  \hline \end{tabular}
\label{tab_strain} \end{table}

Although the qualitative behavior of the bands is very similar for different values of
the strain, topological phase transitions can occur due to some changes
around the Fermi level. In particular, as we show below, the strained
systems I and V in table~\ref{tab_strain} have different topological characters; since the
four-fold band touchings along the tetragonal axis, i.e., the original Dirac
points, are always present in all strained systems, the changing of the
topology must occur through an intermediate phase, in which energy gaps must close somewhere
else in the BZ.

\begin{figure}[h]
\centering{\includegraphics[width=0.46\textwidth]{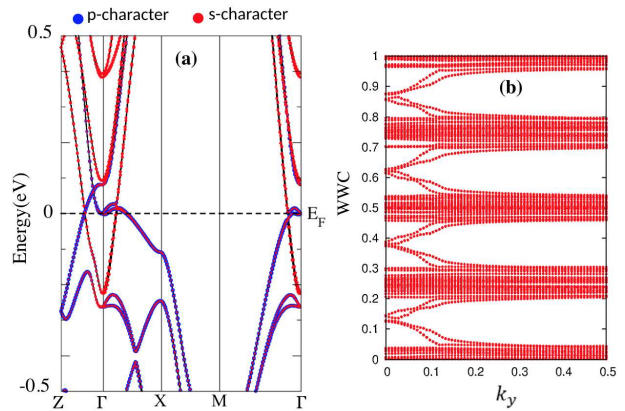}} \caption{(a)
Band structure of the strained system I (see Tab.~\ref{tab_strain}) showing the
Dirac node on the $Z-\Gamma$ high-symmetry line. (b) Evolution of the WCCs in
the $k_z=0$ TRIP, demonstrating the trivial topology of this plane.}
\label{Strain_band} \end{figure}

In Fig.~\ref{Strain_band} the band structure of ``strained system I'' is plotted
along the Z-$\Gamma$-X-M-$\Gamma$ path. This strain corresponds to the case
where the lattice constants $a$ and $b$ are equal to the value of the {\it
relaxed pure} $Cd_3As_2$ (which we therefore refer to as {\it unrelaxed} for the
doped system). Away from the Fermi level, the bands are similar to the fully
relaxed system. Close to the Fermi energy we can see that, besides the usual
four-fold band touching along Z-$\Gamma$, some $k$-points in the $k_z = 0$ plane
along the $\Gamma-X$ and $\Gamma-M$ path present what seems to be band
touchings. A more detailed analysis reveals that these are actually gapped
points, with a small gap of $\Delta E \sim 0.5$ meV along $M-\Gamma$, even
smaller than the one present in the total relaxed system. Furthermore, the node
search with WannierTools did not find any new nodes in the BZ. The $Z_2$
calculations on the $k_z = 0$ TRIP, whose WCC evolution is shown in
Fig.~\ref{Strain_band}b, reveals that the system has the same trivial topology
of the fully relaxed system. This is because a generic horizontal line cuts the
WCC branches in an even number (two) or zero times.

\begin{figure}[h]
\centering{\includegraphics[width=0.46\textwidth]{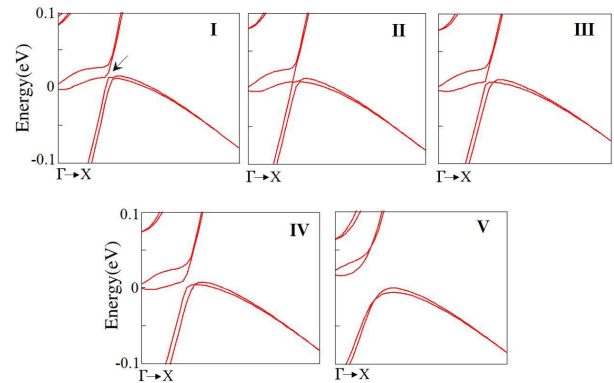}} \caption{Energy
bands along $\Gamma-X$ in a region close to $\Gamma$, which shows how the gap
between the conduction and the valence bands changes with the strain. The
different panels I-V refer to the five strain realizations of Table
~\ref{tab_strain}. The arrow in the panel I marks the point of closest approach of the
conduction and valence bands.} \label{GX} \end{figure}

In order to see how these gaps along the $\Gamma$-X change with different
strains, we plot the evolution of the bands along the $\Gamma-X$ zoomed around
the gaps, for all the five strained systems of Tab.~\ref{tab_strain}. As it is
shown in Figs.~\ref{GX}, the gap is getting smaller until the third case, after
which it starts to increase signaling that if there is any band closings, this
must happen at an intermediate value of the strain. The same thing can be shown
for the $\Gamma-M$ direction. Among the strained systems II-IV in
Table~\ref{tab_strain}, the strained system II gives the smallest gap of 0.4 meV
between the conduction and the valence band along the $\Gamma-X$ and $\Gamma-M$
lines. 

The topological analysis shows that the strained systems II-IV are already in
the non-trivial DSM phase of the pure $Cd_3As_2$ system (see
Table~\ref{tab_strain}). Since the topology changes from the trivial semimetal
phase of strain I to the non-trivial DSM phase of strain II, we deduce that a
band gap closing must necessarily take place for an intermediate value of the
strain. Because of the maintained $S_{4z}$ for this doped system, such a node in
the band structure will have three other copies at rotated positions, and we
expect these 4 nodes to be of Weyl type. Being unable to pin-point the exact
value of the strain corresponding to the band closing, we are unable to
conclusively confirm via a topological analysis that the transition between the
two phases indeed takes place via the mixed Weyl+Dirac phase that we have
discovered for the unrelaxed system. However all these features indicate that
this coexisting phase is the most likely occurrence, which is also supported by
the model Hamiltonian discussed in Sec.~\ref{DWSM}.

\begin{figure}[h]
\centering{\includegraphics[width=0.32\textwidth]{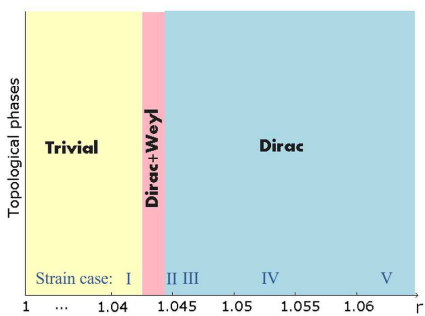}} \caption{Phase
diagram of the topological evolution of the system, in terms of the ratio $ r =
a_{\rm strained}/a_{\rm relaxed}$. The parameters $a_{\rm strained}$ and $a_{\rm relaxed}$ are
the $a$-lattice constants of the totally relaxed ($a_{\rm relaxed}$ =
16.567031 bohr) and of a strained system, respectively.} \label{phase}
\end{figure}

A phase diagram summarizing the topological evolution of $(Cd_{1-x}Zn_x)_3As_2$
($x=0.5$) as a function of applied strain is shown in Fig.~\ref{phase}. Here the
value of the applied strain is expressed in terms of the ratio $r = a_{\rm
strained}/a_{\rm relaxed}$, which tells us how much the first lattice parameter
of the strained cell ($a_{\rm strained}$) is enlarged relatively to the one of
the totally relaxed system ($a_{\rm relaxed}$). 

Summarizing the results for the $(Cd_{1-x}Zn_x)_3As_2$ (x=0.5) system, we found
that  this doped system that breaks IS but maintains both the $S_{4z}$ symmetry
and TRS can exist in three distinct topological phases: (i) a topologically
trivial semimetal phase, still possessing two four-fold nodal points on the
$\Gamma-Z$ axis; (ii) a topological nontrivial DSM phase, characteristic of the
pure system; (iii) a novel mixed Weyl+Dirac phase, first found in the unrelaxed
$(Cd_{1-x}Zn_x)_3As_2$ system, where lattice constants and atomic positions of
pure $Cd_3As_2$ were used. This mixed Weyl+Dirac phase also appears at the
topological phase transition point between the trivial and DSM phase, induced by
an applied external strain, as shown in Fig.~\ref{phase}.

At this point we are not able to conclusively assess from the DFT calculations whether or not
the mixed phase can exist for a finite albeit small range of the parameters that
we use to model the strain in the system. For the systems considered in Table
~\ref{tab_strain}, the value of $c/a$ is kept fixed and equal to the value of
the relaxed doped $(Cd_{1-x}Zn_x)_3As_2$. Changing this parameter can allow
further flexibility to induce nodal points leading to a mixed Weyl+Dirac phase.
Indeed, the  model shows that despite changing some of the parameters in the
system, the mixed phase can still be maintained.

\subsection{Magnetic impurity in $Cd_3As_2$}

Magnetic doping of topological DSMs is of great interest for both fundamental and application reasons,
since the magnetic Weyl phase that can arise due to the broken TRS may support
topological phenomena such as the quantum anomalous Hall effect. In this section
we discuss the effects of magnetic doping in $Cd_3 As_2$. We have investigated
two cases: i) TRS is broken but IS is preserved ii) both TRS and IS are
broken. In both cases, we obtain Weyl phases, which are discussed below. The DFT
calculations are performed using both GGA and GGA+U with $U_{\rm eff}$=4 eV, but
here we mostly discuss the results when $U$ is included.

The stability of $Cd_3 As_2$ doped with different transition metal magnetic
impurities has been investigated and confirmed recently\cite{Jin2015} by calculating the formation energy.
In our study
we have used Mn-doped $Cd_3 As_2$ since Mn  has largest magnetic moment. To preserve
IS, we have introduced two Manganese (Mn) atoms at two Cd sites that are related
by inversion in a 40-atom cell, corresponding to $\sim$ 8\% impurity
concentration. The symmetry of the system reduces to $P2_1/m (C^2_{2h})$, which
contains a mirror plane, $\sigma_h$ and a $C_2$ rotational symmetry along with
inversion. This structure is the same as the one with two Zn impurity discussed
in Sec~\ref{Z2phase}, except that TRS is broken in the present case due to the
magnetic impurity.   

\begin{figure}[h]
\centering{\includegraphics[width=0.46\textwidth]{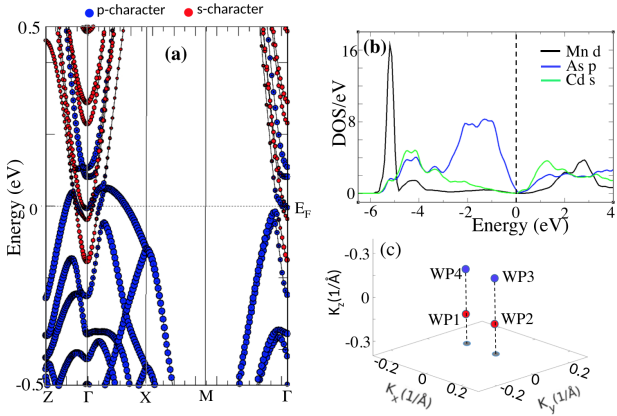}} \caption{(a)
Bandstructure of Mn-doped $Cd_3 As_2$. The bands around the Fermi energy are predominantly of 
As $p$ (blue) and Cd $s$ (red) characters. (b) Projected DOS showing the contribution of
different atoms in the system.  (c) Position of the four Weyl nodes (WP1-WP4), emerging as
a result of broken
TRS. The red (blue) dots represent nodes with
+1 (-1) chirality.}
\label{Mnbands} 
\end{figure}

Fig.~\ref{Mnbands}a shows the bandstructure of Mn-doped $Cd_3As_2$. The correlation among Mn $d$
electrons pushes the occupied $d$ states 5 eV below the Fermi level, as evident from 
the density of states (DOS) shown in Fig.~\ref{Mnbands}b.  Since the correlation strongly localizes the $d$ electrons
of Mn, the hybridization with As $p$ states becomes negligible. Therefore, the
electronic properties around the Fermi level are essentially determined by As
$p$ and Cd $s$ states.  It is also evident from the bandstructure that close to
the Fermi level As $p$ states are higher in energy compared to Cd $s$ states
around the $\Gamma$ point, as in the case of pure $Cd_3As_2$ (Fig~\ref{NoC4band}a),
indicating that band inversion is still preserved in the presence of Mn
impurities. The magnetic moment of Mn atoms are 4.5 $\mu_B$, which is 0.5 $\mu_B$
larger than the corresponding GGA value. Although the $d$ states are absent at
the Fermi level, the magnetization associated with these states play a crucial
role in the topological properties by lifting the spin degeneracies of the bands.

Since the TRS is broken, the double degeneracies of the bands are lifted.
Furthermore, the lack of screw symmetry also breaks the degeneracy along the
tetragonal axis. Consequently, the original four-fold degenerate Dirac points
along $Z-\Gamma$ line disappear. But the presence of Mn impurity introduces a
splitting of the bands which may give rise to new crossing points and
consequently new nodes at generic $\bf k$ points. Therefore, we have searched for the
the nodes in the full BZ, and have found four nodes just 8 meV below the Fermi
level. The position of the nodes are close to the Dirac point of pure $Cd_3
As_2$ on the $k_y$=0 plane with two at ($\pm$ 0.057, 0.0, -0.076)(1/\AA) (WP1, WP2) 
and the other two at ($\pm$ 0.057, 0.0, -0.076)(1/\AA) (WP3, WP4), as shown in 
Fig.~\ref{Mnbands}c. The calculation of the chirality 
shows that WP1 and WP2 have +1 chirality (red) whereas WP3 and WP4 have -1
chirality (blue). Therefore, these nodes are Weyl nodes. The result is
consistent with the expectation that breaking of IS or TRS leads to the
transition to a WSM phase. 

It is interesting to note that unlike the Weyl phase obtained by non-magnetic
impurity (Sec~\ref{weyl_phase}), in this case the two Weyl points below (above)
have the same chirality, which can be understood from the symmetry properties of
the Berry curvature, $\bf{\Omega}(\bf{k})$. As mentioned above, this doped
system contains both inversion and $\sigma_h$ mirror plane. Under IS,
$\bf{\Omega}(\bf{k})$ is even i.e. $I:
\bf{\Omega}(-\bf{k})=\bf{\Omega}(\bf{k})$, and for $\sigma_h$ plane only the
$x$ and $y$ components  of $\bf{\Omega}(\bf{k})$ change sign i.e. $\sigma_h: (\Omega_x,
\Omega_y, \Omega_z) (k_x, k_y, -k_z)=(-\Omega_x, -\Omega_y, \Omega_z) (k_x, k_y,
k_z)$. Consequently, since WP1 and WP3 are inversion symmetric partners, they
have opposite chiralities. Also, since WP1 (WP2) and WP4 (WP3) are related by
$\sigma_h$, they must also have opposite chirality. 

Because of broken TRS, the topological properties of the crystal planes cannot be
characterized by the $Z_2$ invariant. Instead, we have calculated the Chern number
for the planes using WCC methods, which is similar to the calculation of $Z_2$
invariant, except that in this case we use a full path in momentum plane. Since
the nodes lie on the $k_y$=0 plane, we have calculated the Chern number for
$k_x=0$ and $k_z=0$ planes. Our calculation show that the Chern number is 1 only
for the $k_x=0$ plane, which signifies that this plane may be viewed as a 2D Chern
insulator. Therefore, we expect this particular system to support the Quantum
Anomalous Hall Effect, as discussed in section~\ref{intro}.

Finally, we break both TRS and IS by substituting only one Cd by one Mn impurity.
Symmetry is further reduced to $C_s$ point group, which contains only a
$\sigma_h$ mirror plane. The bandstructure is very similar to the one for 2 Mn
impurity shown in Fig.~\ref{Mnbands}a except that a few additional levels appear
due to the reduced symmetry. As before we have searched for nodes in the
BZ, and as expected we have found no Dirac nodes. Instead, we have found four
Weyl nodes, two at $\bf k$-points (0, $\pm$ 0.047, $\mp$ 0.083)(1/\AA)
at 35 meV above the Fermi level with opposite chirality, and two nodes at 
$\bf k$-points (0, $\pm$ 0.056, $\mp$ 0.039)(1/\AA) at 15 meV below the Fermi level, 
also with opposite chirality.
While this system may be viewed as a WSM, it may not be a very promising system
for practical purposes, since the nodes are rather far away from the Fermi level.

We also would like to remark that, in principle, one might think of inducing a coexisting phase 
by breaking TRS while preserving IS, $S_{4z}$ the and mirror symmetries. However, such a realization
requires a high concentration of Mn doping, which is likely to turn the
system into a magnetic metal with trivial topological properties. 

A concise summary of all the results for doped $Cd_3As_2$ obtained in this work
is presented in Table~\ref{tab_sum}.

\begin{table}[h] \caption{Summary of the results for different topological
phases induced by different impurity-doping realizations in $Cd_3As_2$.} \centering
\vspace{1 mm} \begin{tabular}{|c|c|l|}
\hline Impurity         & Symmetry       & Topological properties
\\  \hline		0                & TRS + IS       & DSM phase
\\ & + C$_4$        & Two DPs on the $k_z$ axis.          \\ &                &
$Z_2$=1 only at $k_z$ = 0 TRIP.     \\  \hline 2 Zn             & TRS + IS
& Topological insulator phase          \\ (9\%)		     &                &
A gap of 15 meV at the original DPs. \\ &                & $Z_2$=1 for $k_x$,
$k_y$, $k_z$ = 0 TRIPs. \\ \hline 1 Zn             &   TRS          & WSM phase
\\ (4\%)            &                & 2 DPs splits into 4 Weyl nodes       \\ &
& $Z_2$=1 only at $k_z=0$ TRIP         \\  \hline 12 Zn            & TRS + C$_4$
& Coexisting DSM and WSM phase.        \\ (Unrelaxed)      &                &
Two Dirac nodes on k$_z$ and         \\ (50\%)           &                &
eight Weyl nodes $k_z=0$ plane.      \\  \hline 12 Zn            & TRS + C$_4$
& Trivial semi-metal phase             \\ (Relaxed)        &                &
Two Dirac nodes on $k_z$             \\  \hline 2 Mn             & IS
& Magnetic WSM phase                   \\  (8\%)            &                & 4
Weyl nodes, two with chirality +1  \\ &                & (-1) below (above)
$k_z=0$ plane     \\ \hline 1 Mn             & All broken     & Magnetic WSM
phase                   \\  (4\%)            &                & 4 Weyl nodes on
$k_x=0$ plane but    \\ &                & not related by IS or TRS.
\\ \hline \end{tabular} \label{tab_sum} \end{table}

\section{Topological phase transitions in $Na_3Bi$ DSM} \label{NaBi_DSM}

\begin{figure}[h]
\centering{\includegraphics[width=0.48\textwidth]{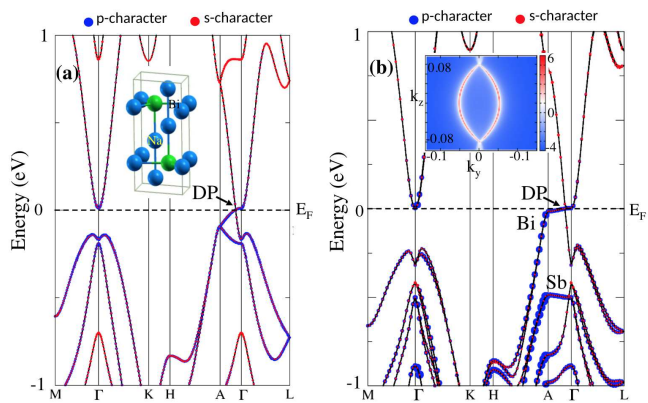}} \caption{(a)
The bandstructure of $Na_3Bi$ DSM indicating the position of the Dirac point on the
$k_z$ axis along the $A-\Gamma$ path. The projections of the Bi, Sb $p$ and Na $s$ states 
are shown in blue and red, respectively. The inset shows the unit cell of $Na_3Bi$. 
(b) The bandstructure of $Na_3Bi$ doped with a single Sb impurity in the unit cell, 
breaking both IS and $S_{6z}$. The Fermi arc surface
states on the [010] surface are shown in the inset.} 
\label{NaBi} 
\end{figure}

In this section we turn our attention to the topological properties of Sb doped
$Na_3Bi$ DSM. Pure $Na_3Bi$ belongs to the hexagonal $P6_3/mmc$ ($D_{6h}^4$)
space group, which is characterize by IS, $6_3$ screw symmetry $S_{6z}$ along
with C$_3$ symmetry. As we will see below, the C$_3$ symmetry has important
consequences in preserving the degeneracy of the bands when the $S_{6z}$
symmetry is lifted by doping. To investigate how doping modifies the topological properties,
we first calculated the electronic structure and topology
of pure $Na_3Bi$. After relaxation, the lattice constants increases from the
experimental value of $a$=$b$=10.29 and $c$=18.25 bohr to
$a$=$b$=10.34 and $c$=18.38 bohr. The bandstructure of this relaxed
system, plotted in Fig~\ref{NaBi}a, shows that the two-fold degenerate valence
band crosses with conduction band on the $k_z$ axis (A-$\Gamma$ path) and forms
a Dirac node, in agreement with previous work\cite{Wang2012}. As in the case of
$Cd_3As_2$, there are two Dirac nodes symmetrically placed around the $\Gamma$
point on the hexagonal axis at (0, 0, $\pm$ 0.086) (1/\AA) with zero chirality. 

To study the effect of non-magnetic impurities on the topological properties, we
have substituted one Bi atom with one Sb, which is isoelectric to Bi. Since there
are only two Bi atoms in the unit cell, this doping corresponds to 50\% impurity
concentration. The space group of the system, consequently, changes to
$P\bar{6}m2$ (or $D_{3h}^1$), which lacks IS but contains a Mirror plane
M$_{110}$ perpendicular to (110) direction as well as the $C_3$ rotational symmetry.
Unlike pure $Na_3Bi$, the relaxation of the doped structure reduces the cell
parameters to $a$ = $b$ = 10.23 and $c$ = 18.21 bohr compared to the
experimental cell parameters of pure $Na_3Bi$ mentioned above. The bandstructure
in Fig.~\ref{NaBi}(b) shows that the $p$ states of Sb lies about 0.5 eV below
the Bi $p$ states. Therefore, the electronic properties around the Fermi level are
essentially determined by Bi $p$ and Na $s$ states. It is also evident that the
band inversion is still preserved along $A-\Gamma$ line as Na $s$ states lies
below Bi $p$ states. However, we note that the band crossing along the
$A-\Gamma$ line is much flatter compared to the pure $Na_3Bi$, suggesting that
at this concentration the system is close to reverting the band inversion.

Since IS is broken, the double degeneracy of the bands are lifted, in general.
However, the anti-commutation of the symmetry operators $M_{110}$ and $C_3$ of
this point group ensures the double degeneracy along the hexagonal axis,
A-$\Gamma$-A line (0,0,$k_z$) (see discussion in  Sec.~\ref{DWSM}). Because of
the maintained band inversion, the doubly degenerate valence and conduction bands
cross on this line at the Fermi energy at (0, 0, $\pm$ 0.083) (1/\AA) forming
two Dirac nodes. The topological analysis confirms that the chirality of the
nodes is zero. As in the case of pure $Na_3Bi$, the $Z_2$  invariant of $k_z=0$
plane is 1. To complete the analysis, we have calculated Fermi arc surface
states as shown in the inset of Fig~\ref{NaBi}(b), which are similar to the arcs
of the pure system. Therefore, our analysis shows that this system with 50\%
impurity is still a DSM. However, we would like to emphasize that although this
phase is considered as a DSM phase because of the presence of two
doubly-degenerate Dirac nodes, the dispersion around the DPs is different from
that of the pure $Na_3Bi$ for the same reason as explained in Sec.~\ref{DWSM}.

When both Bi atoms are replaced by Sb atoms to construct the $Na_3Sb$ crystal, a
gap opens up at the Fermi level. A trivial $Z_2$ invariant confirms that the
system is a trivial insulator. The results shown in this section are consistent
with previous work done using CPA approximation\cite{Narayan2014}.%
   
We have also carried out a study of the effect of broken TRS in
$Na_3Bi$ by substituting one Bi by one Mn atom.  Because of the high Mn
concentration, the system simply turns into a trivial magnetic metal. To address systems
with a lower concentration, it is necessary
to consider considerably larger supercell than the one considered here. 
This in turns makes the topological analysis unwieldy due the need of 
considering a large number of maximally localized
Wannier functions,
as explained in
Sec.~\ref{computation}.

\section{Conclusions} \label{conclusions}

In this work we have used first-principles DFT methods to study the electronic,
magnetic and topological properties of doped DSMs.
Specifically, we have considered $Cd_3As_2$ and $Na_3Bi$, two well-established DSMs, 
focusing in particular on the first, with the goal of investigating
whether it is possible to trigger topological phase transitions by breaking selectively
different symmetries, namely IS, TRS and the rotational symmetry by non-magnetic
and magnetic chemical doping. 

We found that when the rotational (screw) symmetry of $Cd_3As_2$ is broken while
preserving IS, a gap opens up between the conduction and valence
bands, confirming the expectation that the rotational symmetry is crucial for
the stability of the Dirac points. Although the system is still a semimetal due
to the presence of electron and hole pockets, the valence band is completely
separated from the conduction band everywhere in the BZ (see Fig~\ref{NoC4band})
and the $Z_2$ calculation shows that the $k_x, k_y, k_z = 0$ TRIPs are all
topologically nontrivial. If both IS and rotational symmetries are broken, then
each Dirac node splits into two Weyl nodes with opposite chirality i.e. the
system makes a transition to Weyl phase. 

The most interesting topological phase in $Cd_3As_2$ arises when IS is broken
while the rotational screw symmetry is preserved. Due to the presence of the
crystal symmetries such as the mirror plane and the rotational screw axis, along
with the TRS, the two-fold degeneracy of the bands is still maintained along the
tetragonal axis. Therefore, the original Dirac points survive despite the broken
IS. At generic points away from the tetragonal axis, double degeneracy is lifted
and our calculation shows the occurrence of band crossings at eight different
points in the $k_z = 0$ plane, close to the Fermi level. The topological
analysis confirmed that they form four pairs of Weyl nodes, each pair consisting
of nodes with opposite chirality. Therefore, this particular realization of
impurity doping results in a coexisting  Dirac-Weyl phase. 
This numerical finding was corroborated by a continuum ${\bf k} \cdot {\bf p}$ 
which allowed us to elucidate the topological nature of this phase and its stability.    
Although the coexisting phase
was explicitly obtained for an unrelaxed configuration,  we showed that in the
presence of an additional external strain such a phase should emerge at the
strain-induced topological phase transition point separating a trivial semimetal
phase from a DSM phase. It is also possible that this mixed phase could be
present in a system for a small range of external stress around this
transition point.  

We have also investigated the consequence of breaking TRS while preserving IS by
introducing two Mn impurity. Since TRS is broken, the system cannot host any
Dirac node. Instead, the system makes a transition to a magnetic WSM phase with
four Weyl nodes close to the Fermi level. The topological analysis shows that
the Chern number for $k_x = 0$ plane is 1, which signifies that this plane may
be viewed as 2D Chern insulator. Therefore, this magnetically-doped DSM that
transforms into  WSM provides a possible platform to realize the quantum
anomalous Hall effect. In the case when both IS and TRS are broken by a single
Mn impurity in $Cd_3As_2$, it also transform into a magnetic WSM but the Weyl
points reside further away from the Fermi level. 

One final important remark regards the experimental realization of the different
doping cases that we have analyzed in this work. We recognize that achieving the
precise breaking of one particular symmetry while preserving others in a bulk
system is a highly nontrivial endeavor from the experimental point of view.
A more realistic treatment of doping should include some degree of disorder, 
which can possibly affect some of the phases considered here.
However, we believe that some of the cases that we have investigated in this
work, such as the one with a single impurity (both magnetic and nonmagnetic) can
be realized experimentally, since for a large enough unit cell this essentially
represents a random doping. We also believe that the impurity case displaying
the coexisting Dirac+Weyl phase (IS broken but rotational symmetry preserved) can be
realized by delta-doping $Cd_3As_2$ with MBE techniques, forming a layered structure 
(Fig.~\ref{CdAs_struct}) that should be relatively more robust against disorder by individual impurities.
Therefore, we hope that our work will encourage experimental investigations of
impurity-induced phase transitions in these topological semimetal materials.\\   
        
\section*{Acknowledgments} We would like to thank Awadhesh Narayan
for useful comments on Ref.~\onlinecite{Narayan2014}. This work was
supported by the Swedish Research Council (VR) through Grant No. 621-2014-4785,
Grant No. 2017-04404, and by the Carl Tryggers Stiftelse through Grant No. CTS
14:178. Computational resources have been provided by the Lunarc Center for
Scientific and Technical Computing at Lund University. We also acknowledge
CINECA for computer resources allocated under ISCRA initiative and R. Colnaghi
for technical support on computer hardware.

\bibliography{REF}

\end{document}